\documentclass[
superscriptaddress,
 amsmath,amssymb,
 aps, physrev,
twocolumn]{revtex4-2}

\usepackage{graphicx}
\usepackage{dcolumn}
\usepackage{bm}

\begin{document}

\preprint{APS/123-QED}

\title{\textbf{Trimeron ordering, bandgap and polaron hopping in magnetite} 
}%

\author{Nikita Fominykh}
 \email{fominykh.na@phystech.edu}
\affiliation{Joint Institute for High Temperatures of Russian Academy of Sciences, Izhorskaya st. 13 bldg. 2, Moscow, 125412, Russia}
\affiliation{Moscow Center for Advanced Studies, Kulakova str. 20, Moscow, 123592, Russia} 
\author{Vladimir Stegailov}%
 \email{stegailov@jiht.ru}
\affiliation{Joint Institute for High Temperatures of Russian Academy of Sciences, Izhorskaya st. 13 bldg. 2, Moscow, 125412, Russia}
\affiliation{Moscow Center for Advanced Studies, Kulakova str. 20, Moscow, 123592, Russia} 
\affiliation{HSE University, Myasnitskaya Ulitsa 20, Moscow 101000, Russia}

\date{\today}

\begin{abstract}
In this work, we apply the DFT+U approach for a detailed ab initio study of the refined structure of the low-temperature phase of magnetite [M. S. Senn et al., Nature {\bf 481}, 173 (2012)]. We compare the electronic properties of this structure and several alternatives with respect to the presence of trimeron ordering and the bandgap properties. The connection of the trimeron arrangement with site-selective doping of magnetite is discussed. Calculations of the polaron hopping energy allow us to make one step forward toward understanding the complex interplay of polaronic and bandgap contributions to electronic properties of the magnetite.
\end{abstract}

\maketitle


\section{Introduction}

The physics of magnetite has been an enigmatic puzzle for nearly a century~\cite{weiss1914anfangliche,ParksKelly1926} with its pinnacle represented by the Verwey transition~\cite{verwey1939electronic,verwey1941electronic} which is one of the fundamental problems of condensed matter physics and involves a complex interplay of crystal structure and lattice vibrations with electronic structure and strong electron correlations~\cite{walz2002verwey,garcia2004verwey}. The physics of the Verwey transition illustrates the modern level of consistency among experimental methods, theoretical approaches, and computational tools available for understanding the properties of materials. The significant confusion associated with magnetite physics stems from the fact that magnetite was the first substance where the metal-insulator transition was detected: in fact, as a generalization of the Verwey transition. At the same time, modern experimental studies describe this transition undoubtedly as a semiconductor-semiconductor transition (see, e.g.~\cite{schrupp2005high,prozorov2023response}). 

The development of ab initio electronic structure methods follows several complementary avenues~\cite{marzari2021electronic,Bosoni-etal-NatPhys-2024} and materials with strong electronic correlations represent one of the most complicated classes that has progressed during the last decades through many stages of rethinking (see, e.g.,~\cite{tran2006hybrid,chiarotti2024energies,carta2024explicit}). Subtle nuances of electronic structure methods were shown to be crucial for understanding magnetite properties~\cite{rowan2009hybrid,liu2017band,baldini2020discovery,piekarz2021trimeron}.

The question of the crystal structure of magnetite imposes fundamental limitations on the theoretical understanding of its electronic structure. On the one hand, high-temperature (HT) phase has a well established cubic $Fd\bar{3}m$ lattice. However, this lattice is not stable at low temperatures and therefore can not be applied for static zero-temperature ab initio electronic structure calculations. On the other hand, the crystal structure of the low-temperature (LT) phase has been the subject of debate for a long time. Presumably, the first attempt to apply the ab initio theory taking into account strong correlations was made in 1996 by Anisimov~et~al.~\cite{anisimov1996charge} using the DFT+U method. Szoteck~et~al.~\cite{szotek2003ab} applied the SIC-DFT approach for strong correlations and considered cubic, $Pmc2_1$ and $P2/c$ candidate structures. After the next major refinement of the LT phase structure in 2001-2002 by Wright, Attfield, and Radaelli~\cite{wright2001long,wright2002charge}, Leonov~et~al.~\cite{leonov2004charge} and Jeng~et~al.~\cite{jeng2004charge} presented the DFT+U calculations for the $P2/c$ structure. It was shown~\cite{leonov2004charge,zhou2010first} that the resulting orbital ordering obeys the Kugel-Khomskii model~\cite{kugel1982jahn,igoshev2023multiorbital,valiulin2023resistance}.

Subsequent refinement of the LT phase structure was published in 2012 by Senn, Wright, and Attfield~\cite{senn2012charge} showing that the $Cc$ structure was the best variant. Electronic structure calculations for the $Cc$ structure~\cite{senn2012electronic,patterson2014hybrid,liu2017biaxial,srivastava2023density} provided systematically higher bandgap values than predicted by previous studies. To our knowledge, these systematic changes in bandgap values have not received any appropriate attention yet that would be very desirable in the context of the significant complexity of interpretations proposed for high-precision experimental data on magnetite electronic properties~\cite{park1998charge,gasparov2000infrared,baldini2020discovery,piekarz2021trimeron,schrupp2005high,yu2014verwey,hevroni2016tracking,prozorov2023response}.

The key feature of the Verwey transition is a sharp decrease in conductivity about two orders of magnitude below $T_V\sim125~\text{K}$. It was this transition that stimulated the concept of metal-insulator transition~\cite{mott1968metal} frequently used to describe the differences between the LT and HT phases of magnetite. The closure of the bandgap at temperatures above $T_V$ is usually implied. Infrared optical conductivity data~\cite{park1998charge,gasparov2000infrared}, the photoemission spectroscopy (PES) results~\cite{bishop1974photoemission,shiratori1986photoemission,chainani1995high,wang2013fe,taguchi2015temperature,schmitt2021bulk} and scanning tunneling spectroscopy (STS) results~\cite{shimizu2010termination,yu2014verwey} could be interpreted as the manifestation of a bad metallic state above $T_V$. However, these methods can be quite sensitive to surface states, and at the same time no clear Fermi edge is observed. Moreover, the conductivity value itself is much lower than the minimum Ioffe-Regel conductivity for metals~\cite{yu2014verwey}. For the HT phase, the dc conductivity follows the Arrhenius law with an activation energy $E_a^{HT}\sim{50}~\text{meV}$. In the LT phase, the conductivity follows the Arrhenius law also with a slightly higher activation energy $E_a^{LT}\sim{100}~\text{meV}$~\cite{verwey1939electronic,matsui1977specific,kuipers1979electrical,prozorov2023response}, similar values are also observed for terahertz conductivity~\cite{pimenov2005terahertz}.

There are several models that describe charge transport in magnetite. The bandlike model with delocalized electron transport proposed by Cullen and Callen~\cite{cullen1970collective}. In contrast, other models suggest localized charge transport via polarons formed because of strong electron-phonon interactions, such as small polarons~\cite{mott1980materials}, bipolarons~\cite{chakraverty1974verwey} and molecular polarons~\cite{yamada1980molecular}. Later, Ihle and Lorenz assumed the superposition of the small polaron band and the hopping conductivity~\cite{ihle1986small}. 
The conductivity behavior shows a clear semiconductor-semiconductor type of Verwey transition~\cite{prozorov2023response}.

The concept of a semiconductor-semiconductor transition is further supported by a series of papers, in which the bandgap $E_{g}$ is observed both above and below $T_V$. Using temperature-dependent high-resolution PES data, Park et al.~\cite{park1997single} showed that the bandgap does not vanish above the Verwey transition but undergoes a jump decrease from $E_{g}^{LT}\sim150$~meV to $E_{g}^{HT}\sim100$~meV. Later, similar results were obtained using two less surface-sensitive methods: the soft X-ray PES~\cite{schrupp2005high} and the hard X-ray PES~\cite{kimura2010polaronic} gave slightly lower values $E_{g}^{LT}\sim100$~meV and $E_{g}^{HT}\sim50$~meV, which are in a good agreement with the conductivity activation energies. In all of these works, the possible role of polarons on electronic properties is emphasized.
In addition, a STS study by Jordan~et~al. is reported in which a constant $E_{g}^{LT}\approx E_{g}^{HT}\sim200$~ meV surface gap is observed below and above the Verwey transition~\cite{jordan2006scanning}. Later, with a single magnetite nanocrystal, Hevroni et al. observed a decrease in the bandgap just above the transition to $E_{g}^{HT}\sim75$~meV~\cite{hevroni2016tracking} and in another study of single magnetite nanocrystal a higher value for low temperature phase $E_{g}^{LT}\sim320$~meV and $E_{g}^{HT}\sim70$~meV were observed~\cite{banerjee2019track}.

\begin{table*}
\centering
\caption{The review of the selected previous ab initio studies of magnetite that provided the bandgap $E_g$ values. Selected key properties of previous ab initio magnetite studies. The Fe$_B$ column show if charge separation was detected in the model. Approaches for strong electron correlations, lattice symmetry groups and maximum numbers of atoms are shown. The comments column contains either the quantity used for comparison with experimental data or a unique aspect of a particular study. }
\label{tab:dft_studies}
\setlength{\tabcolsep}{5pt}
\renewcommand{\arraystretch}{1.4}
\centering
\begin{tabular}{cccccrl}
\hline
$E_g$, eV & Fe$_B$ state & Strong correlations & Lattice &  $N$ & Reference & Comment \\
\hline
\hline
\multicolumn{7}{c}{Cubic high temperature phase} \\
\hline
0 & Fe$_B^{2.5+}$ & - & $Fd\bar{3}m$ & 14 & Yanase and Siratori, 1984~\cite{yanase1984band} & Exp.: $E_g = 0~\text{eV}$~\cite{bishop1974photoemission} \\ 
0 & Fe$_B^{2.5+}$ & - & $Fd\bar{3}m$ & 14 & Zhang and Satpathy, 1991~\cite{zhang1991electron} & Exp.: $E_g = 0~\text{eV}$~\cite{bishop1974photoemission} \\
0 & Fe$_B^{2.5+}$ & - & $Fd\bar{3}m$ & 14 & Yanase and Hamada, 1999~\cite{yanase1999electronic} & Exp.: $E_g = 0~\text{eV}$~\cite{chainani1995high} \\
0 & Fe$_B^{2.5+}$ & - & $Fd\bar{3}m$ & 56 & Hendy et al., 2003~\cite{hendy2003ab} & Point defects \\
0 & Fe$_B^{2.5+}$ & $U=4.0$, $J=0.8$ & $Fd\bar{3}m$ & 56 & Piekarz et al., 2007~\cite{piekarz2007origin} & Exp.: phonon spectra~\cite{samuelsen1974low} \\
0 & Fe$_B^{2.5+}$ & $U_{eff}=3.94$ & $Fd\bar{3}m$ & 56 & Arras et al., 2013~\cite{arras2013electronic} & Cationic defects \\
0 & Fe$_B^{2.5+}$ & $U_{eff}=4.0$ & $Fd\bar{3}m$ & 56 & Noh et al., 2014~\cite{noh2014density} & \\
$\sim$0.4 & Fe$_B^{2+/3+}$ & $HSE06$ & $Fd\bar{3}m$ & 56 & \\
0 & Fe$_B^{2.5+}$ & $HSE(15\%)$ & $Fd\bar{3}m$ & 56 &  & \\
0.18 & Fe$_B^{2+/3+}$ & $U_{eff}=3.5$ & $Fd\bar{3}m$ & 14 & Liu and Di~Valentin, 2017~\cite{liu2017band} & Exp.: $E_g$ = 0.1-0.2~eV\\
0.42 & Fe$_B^{2+/3+}$ & $HSE06$ & $Fd\bar{3}m$ & 14 & & \cite{park1997single,jordan2006scanning}\\ 
0.34 & Fe$_B^{2+/3+}$ & $B3LYP$ & $Fd\bar{3}m$ & 14 \\
0.06 & Fe$_B^{2+/3+}$ & $U_{eff}=3.5$ & $Fd\bar{3}m$ & 448 & Shutikova and Stegailov, 2022~\cite{shutikova2022frenkel} & Exp.: defect energies~\cite{dieckmann1986defects} \\
\hline
\multicolumn{7}{c}{Monoclinic low temperature phase} \\
\hline
0.34 &  Fe$_B^{2+/3+}$ & $U=4.51$, $V=0.18$  & $Imma$ & 14 & Anisimov et al., 1996~\cite{anisimov1996charge} & Exp.: $E_g = 0.14~\text{eV}$~\cite{chainani1995high} \\
0.1 & Fe$_B^{2+/3+}$ & SIC-LSD & $P2/c$ & 56 & Szotek et al., 2003~\cite{szotek2003ab} &  \\
0 & Fe$_B^{2.5+}$ &  - & $P2/c$ & 56 & Leonov et al., 2004~\cite{leonov2004charge} & Exp.: $E_g = 0.14~\text{eV}$~\cite{park1998charge} \\
0.18 & Fe$_B^{2+/3+}$ & $U=5$, $J=1$  & $P2/c$ & 56 \\
0.2 & Fe$_B^{2+/3+}$ & $U=4.5$, $J=0.8$  & $P2/c$ & 56 & Jeng et al., 2004~\cite{jeng2004charge} & Exp.: $E_g = 0.1~\text{eV}$~\cite{park1998charge} \\
0.075 &  & LSDA+DMFT &  &  & Craco et al., 2006~\cite{craco2006verwey} &  \\
0.33 & Fe$_B^{2+/3+}$ & $U_{eff}=5$ & $C2/c$ & 14 & Pinto and Elliot, 2006~\cite{pinto2006mechanism} & Exp.: $E_g = 0.14~\text{eV}$~\cite{park1998charge} \\ 
0.33 & Fe$_B^{2+/3+}$ & $U=4.0$, $J=0.8$ & $ P2/c$ & 56 & Piekarz et al., 2007~\cite{piekarz2007origin} & Exp.: $E_g = 0.14~\text{eV}$ ~\cite{park1998charge} \\
0.87 & Fe$_B^{2+/3+}$ & $B3LYP$ & $P2/c$ & 56 & Rowan et al., 2009~\cite{rowan2009hybrid} & Exp.: Raman modes~\cite{gasparov2000infrared} \\
0.32 & Fe$_B^{2+/3+}$ & $B3LYP(15\%)$ & $P2/c$ & 56 &  &  \\
0.5 & Fe$_B^{2+/3+}$ & $U=4.5$, $J=0.9$ & $Cc$ & 112 & Senn et al., 2012~\cite{senn2012electronic} & Exp.: $E_g = 0.1~\text{eV}$~\cite{gasparov2000infrared} \\
0.5 & Fe$_B^{2+/3+}$ & $B3LYP(10\%)$ & $Cc$ & 112 & Patterson, 2014~\cite{patterson2014hybrid} & Exp.: NMR resonance \cite{mizoguchi2001charge} \\
0.9 & Fe$_B^{2+/3+}$ & $B3LYP(15\%)$ & $Cc$ & 112 &  & \\
0.51 & Fe$_B^{2+/3+}$ & $U=4.5$, $J=0.89$ & P2/c & 112 & Liu et al, 2017~\cite{liu2017biaxial} & Biaxial strain \\
1.0 & Fe$_B^{2+/3+}$ & $U=4.5$, $J=0.89$ & $Cc$ & 112  \\
$\sim$1.0 & Fe$_B^{2+/3+}$ & $U_{eff}=4.1$ & $Cc$ & 448 & Srivastava et al., 2023~\cite{srivastava2023density} & Oxygen vacancies \\
\hline
\end{tabular}
\end{table*}

In the context of such complex experimental data, ab initio calculations could provide a complementary point of view on electronic and optical properties. Despite the fact that strongly correlated iron oxides are very challenging systems for density functional theory calculations~\cite{meng2016density}, a considerable progress has been made in this field. There are cubic $Fd\bar{3}m$ phase calculations that describe magnetite as a half-metal~\cite{yanase1984band,zhang1991electron,yanase1999electronic,hendy2003ab,pinto2006mechanism,piekarz2007origin,arras2013electronic,noh2014density,naveas2023first}. Liu and Di Valentin showed that charge separation leads to a small bandgap in the cubic phase~\cite{liu2017band}. 
Orbital ordering and trimerons structural properties in the LT phase have also been thoroughly studied~\cite{anisimov1996charge,leonov2004charge,yamauchi2009ferroelectricity,zhou2010first,yamauchi2012orbital,senn2012electronic,liu2017biaxial,srivastava2023density}. Furthermore, nuclear magnetic resonance (NMR) and M\"{o}ssbauer spectrs were successfully explained within the framework of ab initio calculations~\cite{patterson2014hybrid,vreznivcek2015hyperfine,vreznivcek2017understanding}. Later, within the DFT+U framework, the Frenkel pair formation energy and the bandgap together were described in the cubic $Fd\bar{3}m$ phase~\cite{shutikova2022frenkel}. The essential results of the selected ab initio studies of HT and LT phases relevant to this work are presented in Table~\ref{tab:dft_studies}.

Recently, polaron excitations and trimeron ordering have attracted a great deal of attention. Resonant inelastic X-ray scattering revealed spin-orbital excitations driven by polaronic distortion that persist even in the HT phase~\cite{huang2017jahn}. Terahertz spectroscopy and ab initio calculations have been used to study the interplay between trimeron order and specific phonon modes~\cite{baldini2020discovery,piekarz2021trimeron}. Spin transport properties in two-dimensional magnetite nanosheets were associated with the Verwey transition~\cite{jia2024spin}. 
Another complementary topic is pressure effects on the Verwey transition~\cite{mori2002metallization,rozenberg2006origin,kozlenko2019magnetic}.
In addition, trimerons and polaronic transport have been examined in a similar iron oxide Fe$_4$O$_5$~\cite{layek2022verwey,ovsyannikov2023electronic}. Significant progress has been also made for hematite Fe$_2$O$_3$ in the description of polaronic transport properties~\cite{ahart2022electron,shelton2022polaronic,redondo2024real} and polaronic excitons~\cite{rassouli2024electronic,rassouli2024excitons}. 

In this paper, we will focus on the LT phase of magnetite and make an attempt to reveal a comprehensive interplay of trimeron, orbital, and charge orderings, considering together the atomic structure, band structure, and polaron hopping in the framework of a DFT+U model. The reminder of the paper is organized as follows. In Section~II, the general model and computational details are given. In Section~III, the orbital-charge orderings together with the band structures and the trimeron orderings are considered. In Section~IV, we present small polaron hopping calculations.
In Section~V, we discuss our results in the context of possible interpretations of polaron and bandgap features detected in optical conductivity data. The paper is concluded in Section~VI.

\begin{figure*}
    \centering    \includegraphics[width=0.98\linewidth]{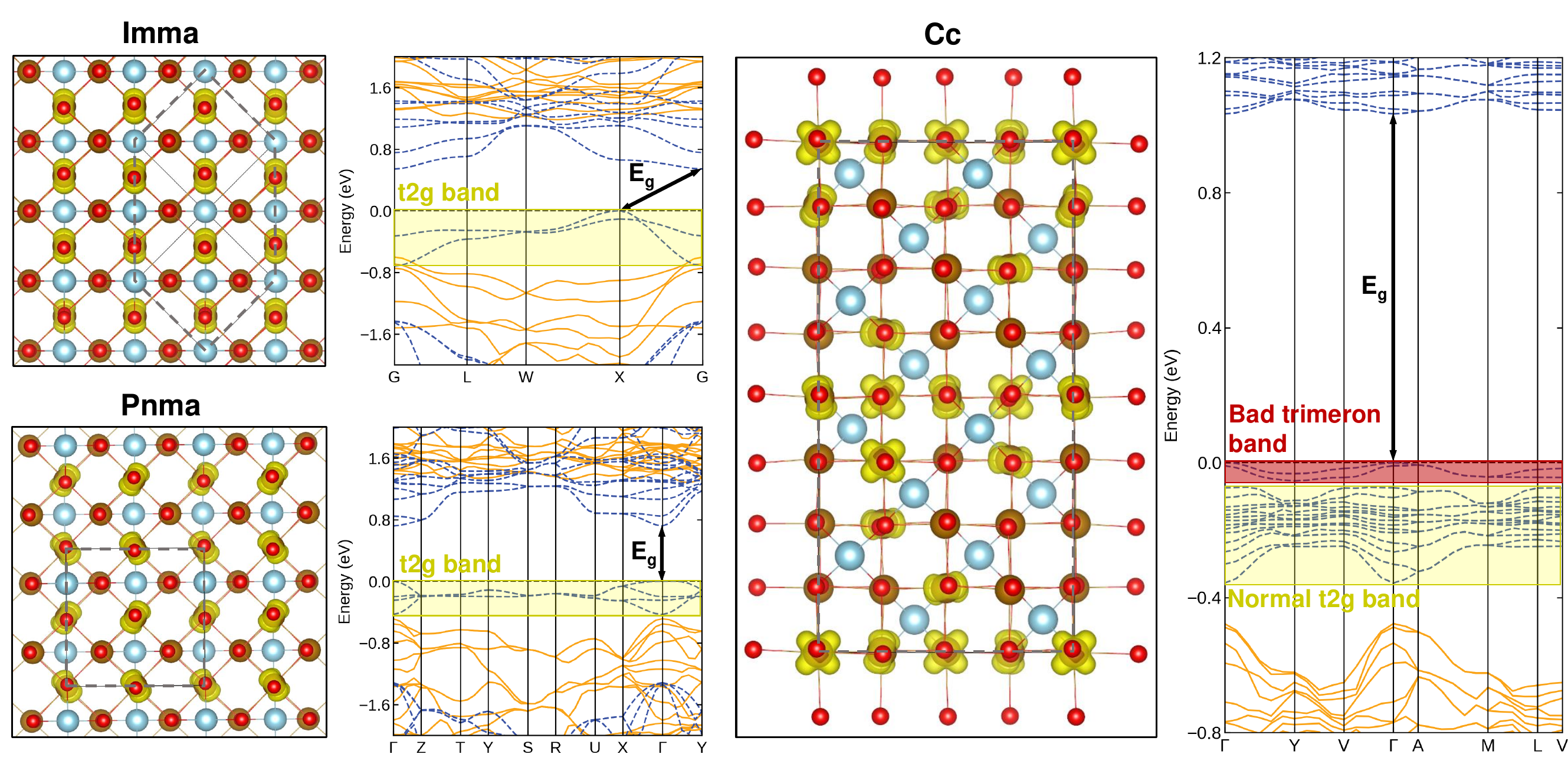}
    \caption{Orbital-charge ordering and band structure. Charge density isosurfaces are visualized in VESTA~\cite{momma2011vesta}. Iron atoms in tetrahedral and octahedral sites are shown in blue and brown, oxygen atoms are red. Gray dashed lines represent periodic boundaries for primitive lattice. Yellow isosurfaces show occupied $t_{2g}$ orbitals of octahedral Fe$^{2+}$ atoms which energy bands lie just below the Fermi level. Band structures are plotted using sumo~\cite{ganose2018sumo}: orange lines show spin-up bands and blue dashed lines show spin-down bands. For the $Cc$ structure the $t_{2g}$ bands forming ``bad'' trimerons are also highlighted in red. }
    \label{fig:orbital}  
\end{figure*}

\begin{table*}
\centering
\caption{DFT+U results obtained for various charge-orbital orderings. Calculations are performed from supercells with an initial symmetry group which contains $N$ atoms. The final symmetry shows the symmetry group obtained after cell and ion relaxations determined by sumo~\cite{ganose2018sumo}. $\Delta E$ represents the difference in energy per one functional unit with respect to the $Cc$ structure~\cite{senn2012charge}. $E_g$ is the bandgap calculated as the difference of the HOMO and LUMO bands energies. The bandgap type is obtained for the primitive cell that corresponds to the  final symmetry. We assume that trimerons are linear 3-site distortions, which, in general, consist of one Fe$^{2+}$ atom in the middle and two Fe$^{3+}$ atoms at the edges. We define the trimeron ordering as some structured arrangement of trimerons (see Figure~\ref{fig:trimerons}).}
\label{tab:orderings}
\setlength{\tabcolsep}{5pt}
\renewcommand{\arraystretch}{1.4}
\centering
\begin{tabular}{ccccccc}
\hline
Initial symmetry & N & Final symmetry & $\Delta E$, meV/f.u. & $E_g$, eV & Bandgap type & Trimeron ordering\\
\hline
\hline
$Imma$  & 14/56   & $Imma$ & $33$ & 0.55 & indirect & no \\
$Imma$  & 112/224 & $Pnma$ & $20$ & 0.72 & direct   & yes \\
$P2/c$  & 112/224 & $P2/c$ & $5$  & 0.86 & indirect & yes \\
$P4_122$& 224 & $P4_12_12$ & $1$  & 1.12 & indirect & no \\
$P4_122$& 112 & $P2_1$  & $<1$  & 1.09 & indirect & no \\
$Cc$    & 112/224 & $Cc$   & $0$  & 1.03 & direct   & yes \\
\hline
\end{tabular}
\end{table*}

\section{Computational details}

Spin-polarized DFT calculations are performed using the projector-augmented wave (PAW) method~\cite{blochl1994projector,kresse1999ultrasoft} implemented in VASP~\cite{kresse1993ab,kresse1996efficient}. The energy
cut-off of the plane wave basis is set to 550 eV. The exchange correlation energy in the generalized gradient approximation by Perdew, Burke, and Ernzerhof (GGA-PBE~\cite{perdew1996generalized}) is used. $\Gamma$-centered k-meshes from $2\times2\times2$ to $8\times8\times8$ with respect to supercell size are used, employing the Bl\"ochl tetrahedron corrections~\cite{blochl1994improved}. The k-mesh convergence obtained is of the order of one meV per functional unit. Optimal strategies are used to parallelize the computations~\cite{stegailov2019vasp}.
Strong electron correlations are taken into account with a Hubbard $U_\text{eff} = U-J$ correction applied to 3d electrons of Fe based on the rotation-invariant Dudarev approach~\cite{dudarev1998electron}. $U_\text{eff}  = 3.8~\text{eV}$ is used that is consistent with previous studies on magnetite (see Table~\ref{tab:dft_studies}) and was used in previous works on cubic phase of magnetite~\cite{liu2017band,shutikova2022frenkel}, on chromite~\cite{fominykh2023polarons} and on nickel ferrite~\cite{fominykh2025influence}. There are some approaches to obtain $U_{\text{eff}}$ entirely ab initio~\cite{cococcioni2005linear,aryasetiawan2006calculations,falletta2022hubbard,falletta2023polaron}, however, a fine-tuned $U_{\text{eff}}$ value for a certain property could be not optimal for taking into account simultaneously all ground state properties of a system (see, e.g.~\cite{franchini2021polarons}).

\begin{figure*}
    \centering  \includegraphics[width=0.995\linewidth]{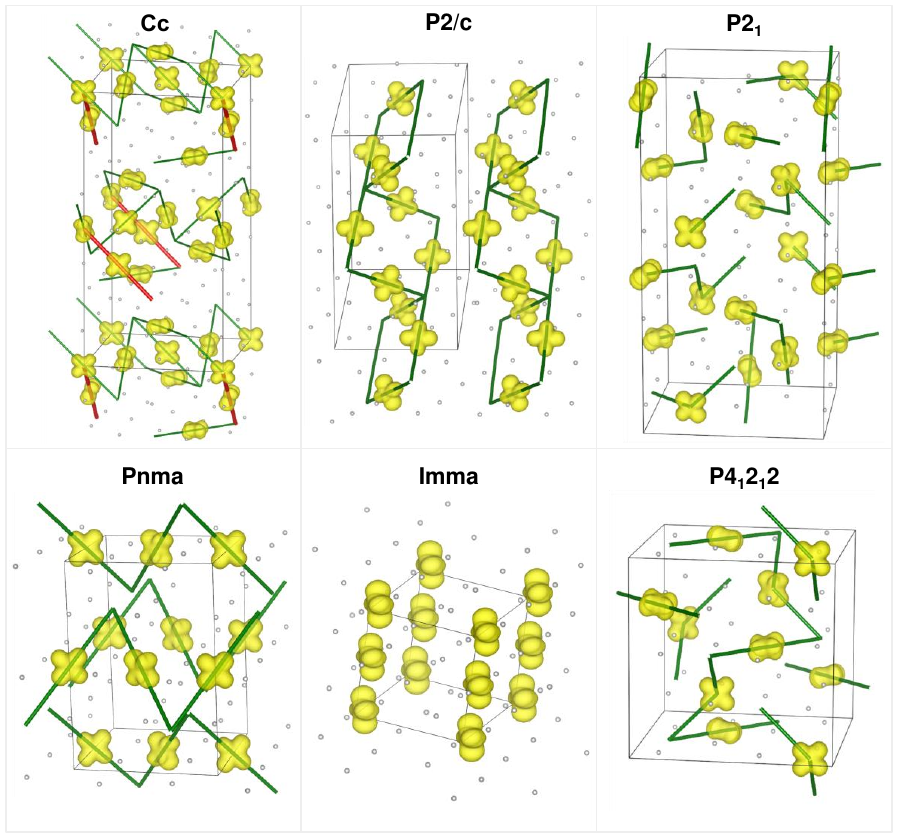}
    \caption{Trimeron ordering in the charge-orbital orderingsconsidered. Yellow isosurfaces show occupied $t_{2g}$ orbitals of octahedral Fe$^{2+}$ atoms. Oxygen atoms are shown in gray, while iron atoms are hidden. Green lines show shortening of Fe$^{2+}$-Fe$^{3+}$ bonds which form trimerons. For the case of the $Cc$ structure, we assume similar trimeron ordering with two long bonds and a ``bad'' trimeron as it was proposed by Senn et~al.~\cite{senn2012charge}. The ``bad'' trimeron is shown as a bold red line. For other cases we show Fe$^{2+}$-Fe$^{3+}$ bonds that are shorter than $2.98$~\AA. Thus, for the orbital-charge orderings with the symmetry group $P2/c$ and $Pnma$ a well identified trimeron order can be seen. In the $Imma$ structure there is no shortening of Fe$^{2+}$-Fe$^{3+}$ bonds and therefore no trimerons, while in the $P2_1$ and $P4_12_12$ structures in addition to linear distortions which could be referred to as trimerons, there is another type of some angle-like distortions.}
    \label{fig:trimerons}  
\end{figure*}

\section{Orbital-charge ordering and trimerons}

In order to study the complex interplay of charge, orbital, and trimeron orderings, we consider various charge orderings within one DFT+U framework.
We consider 4 types of probable initial charge ordering with the following symmetry space groups: the Verwey-like $Imma$ ordering, the experimentally suggested $Cc$ and $P2/c$ orderings, as well as the $P4_122$ ordering predicted for nickel ferrite~\cite{sharma2022influence} possessing a similar structure of B-sublattice of Ni$^{2+}$/Fe$^{3+}$ ions.

We perform our ground-state calculations in the three-stage approach: 
\begin{itemize}
    \item[1.] The initial coordinate relaxation of the distorted cubic phase with a symmetry group of interest to obtain desired Fe$^{2+}$/Fe$^{3+}$ distribution (the structure from~\cite{senn2012charge,pachoud2020site} is used in the case of $Cc$).
    \item[2.] The full coordinates relaxation with the possibility of changing the shape and volume of the supercell.
    \item[3.] The determination of the final symmetry groups and the calculations of the band structure in the corresponding primitive cells.
\end{itemize}

\begin{figure*}
    \centering
    \includegraphics[width=1.0\linewidth]{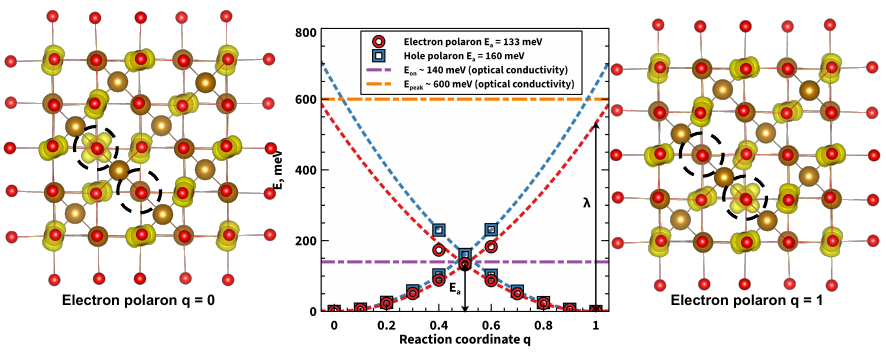}
    \caption{The energy profiles for polarons hopping (in the center) and the electronic structure pictures of an electron polaron localized states (the left and right pictures). The activation energy $E_a$ and the reorganization energy $\lambda$ for electron polaron hopping are marked by arrows. The dash-dotted horizontal lines show the experimental data on the onset energy ($E_\text{on1}$) and the first peak energy ($E_\text{peak}$) from the optical conductivity experiments~\cite{park1998charge,gasparov2000infrared}.}
    \label{fig:polscheme}  
\end{figure*}

The results obtained are provided in Table~\ref{tab:orderings}. In agreement with the experiment, the structure with the $Cc$ symmetry group turned out to be the lowest in energy. Considering the different sizes of the initial supercells in the case of the $Cc$ and $P2/c$ symmetry groups, there are no noticeable differences. Quite interesting results are obtained in the case of relaxation of the $Imma$ and $P4_122$ orderings. So in the case of $Imma$ symmetry group, we see a transition to $Pnma$ symmetry group after relaxation for supercells of 112 and 224 atoms. Furthermore, we obtaine $P2_1$ and $P4_12_12$ symmetry groups after relaxation in the case of 112 and 224 atoms initially with the $P4_122$ symmetry group.
For all six final symmetries, we obtain the quasi-cubic lattice parameter of $\bar{a}_0 = 8.49$~\AA, which agrees with experimental $\bar{a}^{\text{exp}}_0 = 8.39$~\AA  reasonable well.
The difference in the energy of different orderings is rather small, especially in the case of structures with symmetry groups $P4_12_12$ and $P2_1$. This may indicate the existence of inhomogeneous ordering inclusions of this type in crystals at finite temperature.

An intriguing result that should be highlighted is the case of the transition from $Imma$ to $Pnma$ symmetry, in which the charge ordering in the B sublattice does not change, but at the same time the orbital ordering changes and forms chains of trimerons. This clearly shows the importance of understanding not only the charge distribution but also the finer ordering of the orbitals. Moreover, this change in orbital ordering is accompanied by a transition from an indirect to a direct bandgap, which could be seen in the band structures shown in Figure~\ref{fig:orbital}.

After relaxation of our model with $Cc$ ordering, we observe two off-centered distorted trimerons with a longer bond and the so-called ``bad'' trimeron, which is a Fe$^{2+}$-Fe$^{2+}$-Fe$^{3+}$ complex (Figure~\ref{fig:trimerons}). The ``bad'' trimeron is a really fascinating feature, it is one of the main differences with the $P2/c$ symmetry structure which has a bit higher energy and also has some sort of trimeron ordering (Figure~\ref{fig:trimerons}), the indirect bandgap with a bit smaller width (Table~\ref{tab:orderings}).
Moreover, the obtained band structure allows us to distinguish the band with quite small dispersion formed by $t_{2g}$ orbitals on B42 sites (we use a similar notation of non-equivalent sites as~\cite{senn2012charge,senn2012electronic}), which is the constituent of the ``bad'' trimerons. 

A recent study of Zn-doped magnetite~\cite{pachoud2020site} showed selective oxidation of the ``bad'' trimeron B42 site, despite the fact that orbital order is usually suppressed by chemical doping.

Since the ``bad'' trimeron band lie higher in energy than the others, this could be a suitable explanation for the mechanism for the experimentally observed phenomenon. Thus, Zn$^{2+}$ cations randomly substitute Fe$^{3+}$ ions in the A sublattice, which does not distort the charge ordering much, leading to oxidation of iron cations at the B42 site. This experimentally observed ``bad'' trimeron feature further confirms that the ground state of the LT phase has structures with the $Cc$ symmetry group, and can also serve as an indication of the importance of the properties of ``bad'' trimerons in the destruction of the trimeron order at the Verwey transition.

\begin{figure*}
    \centering
    \includegraphics[width=0.55\linewidth]{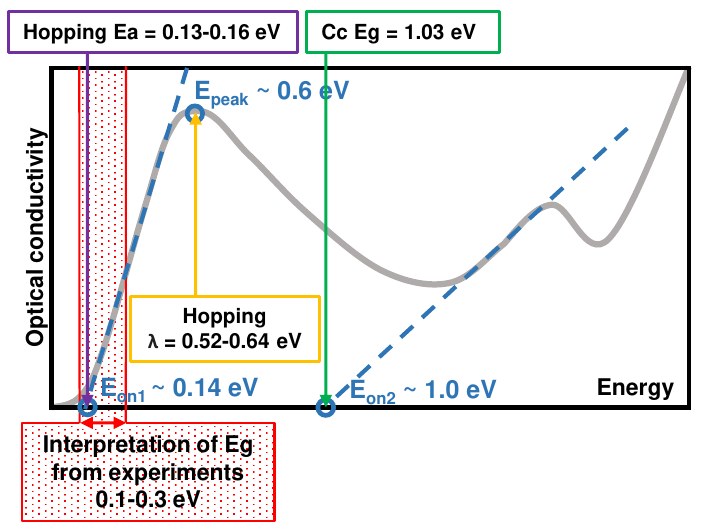}
    \caption{The schematic interpretation of optical conductivity. The blue circles show the onset energies and peak energies from the experiments~\cite{park1998charge,gasparov2000infrared}. The red range shows the bandgap interpretations from various experiments~\cite{chainani1995high,park1997single,park1998charge,gasparov2000infrared,schrupp2005high,jordan2006scanning,shimizu2010termination,kimura2010polaronic,taguchi2015temperature,hevroni2016tracking,banerjee2019track}. The violet, yellow and green boxes show our calculated polaron hopping activation energy $E_a$, polaron reorganization energy $\lambda$ and bandgap $E_g$.}
    \label{fig:parkscheme}  
\end{figure*}

\section{Small polaron hopping}

We threat small polaron hopping within the Holstein-Mott theory~\cite{emin1969studies,holstein2000studies,austin1969polarons}, implementing the Marcus approach \cite{marcus1985electron,marcus1993electron} in our first-principle DFT+U model. We use the linear interpolation of atomic coordinates between two localized states to consider the process of nonadiabatic electron transport by transfer of a small electron or hole polaron representing an excess charge self-trapped by a local lattice distortion (the illustration of a localized electron polaron in the initial and final states is presented in Figure~\ref{fig:polscheme}). This approach was shown to be effective in a number of previous works on titanium dioxide~\cite{deskins2007electron}, hematite~\cite{ahart2022electron}, chromite~\cite{fominykh2023polarons}, and nickel ferrite~\cite{fominykh2025influence}.

In the first step, the localized polaron is obtained by ionic relaxation with a unity change in the number of electrons. To reduce convergence problems, ion relaxation is performed in such a way that only the nearest oxygen atoms are optimized first, and then the full relaxation of all atoms is performed. 
Since electron transfer in the polaron hopping process can occur only if there is sufficient local lattice distortion, the atom coordinates could be a suitable reaction coordinate. In the second step, the energy barrier is obtained from the calculated energy surface along the linearly interpolated coordinates beyond the localized state $\mathbf{R_{q}} = (1-q)\mathbf{R_{q=0}} + q\mathbf{R_{q=1}}$.

There are a large number of possible elementary paths in the $Cc$ ground state caused by a high number of non-equivalent localized polaron sites. Therefore, for the first approximation of the polaron activation energy, we use the simplest $Pnma$ LT structure with trimeron ordering, where we can estimate a path-averaged activation energy $E_a$ with the nearest hopping energy barrier. In this case, we have two possible nearest hopping paths in the octahedral sublattice: these are the electron polaron hopping along Fe$^{3+}$ chains and the hole polaron hopping along Fe$^{2+}$ chains.  

Thus, from the energy profiles (Figure~\ref{fig:polscheme}) we see a completely nonadiabatic behavior of the transfer processes without an explicit transition state. The profiles have a parabolic form and are symmetric with respect to the transition state $q=0.5$. Thus, according to Marcus theory, the activation energies $E_a$ of electron and hole polaron migration are equal to 133 and 160~meV that agree with those of other iron oxides~\cite{deskins2007electron,ahart2022electron,fominykh2023polarons,fominykh2025influence} and provide new insights into the interpretation of the experimentally observed  electronic properties of magnetite.

\section{Discussion}

Trimeron ordering has a significant effect on the ground state orbital and charge ordering. Polaron excitations probably lead to the vanishing of the trimeron ordering above $T_V$. However, our calculations for the $Pnma$ structure show that a single polaron distorts the surrounding trimeron ordering \textit{only slightly}.

In our $Pnma$ model of polaron transport in the LT phase, the activation energies obtained $E_a$ = 0.13-0.16~eV are quite close to the experimentally observed values of $\sim$0.1-0.2~eV from conductivity measurements~\cite{verwey1939electronic,matsui1977specific,kuipers1979electrical,prozorov2023response} and to the calculated values of 87-200 meV obtained within the coherent polaron tunneling model by Baldini et al.~\cite{baldini2020discovery}, which describes low energy electronic collective modes of trimeron ordering. Moreover, these values are well aligned with the characteristic activation energies of diabatic polaron hopping in other iron oxides of 0.09-0.19~eV for Fe$_2$O$_3$~\cite{ahart2022electron}, $0.16~\text{eV}$ for FeCr$_2$O$_4$~\cite{fominykh2023polarons}, and 0.12-0.18~eV for NiFe$_2$O$_4$~\cite{fominykh2025influence}. When considering polaron transport in more detail, together with taking into account all possible elementary transport pathways in the $Cc$ structure, it makes sense to consider adiabatic transport as well. For hematite Ahart et al.~\cite{ahart2022electron} showed that explicit consideration of electron coupling leads to a decrease in activation energies to 0.07-0.08~eV and this reduction is more pronounced in the case of hole polarons.

The complex interplay of polaron transport and interband transitions gives a freedom of interpretation of the electronic structure properties obtained in the experiment. Thus, the early works within the framework of ab initio calculations gave the bandgap values of 0.18-0.34~eV~\cite{anisimov1996charge,leonov2004charge,jeng2004charge,pinto2006mechanism,piekarz2007origin} that seemed to be only slightly overestimated in comparison with the then accepted interpretation of experimental values of 0.1-0.2~eV. 
The use of a more correct structure of the LT phase with the $Cc$ symmetry group should lead to a better agreement with experiment.
However, in recent studies, much larger bandgap values of 0.5-1.0~eV were observed~\cite{senn2012electronic,patterson2014hybrid,liu2017biaxial,srivastava2023density}. Our calculations for various orbital and charge orderings fit well in this range with $E_g$=0.55-1.12~eV (see Table~\ref{tab:orderings}). 

For the most stable structure with the $Cc$ symmetry group, we obtian the bandgap of $1.03~\text{eV}$ that agrees well with the previous calculations~\cite{patterson2014hybrid,liu2017biaxial,srivastava2023density}. The discrepancy with the value of $0.5~\text{eV}$ obtained by Senn et al.~\cite{senn2012electronic} is presumably due to the absence of relaxation of the structure in that study.
Thus, we conclude that the bandgap in the LT phase is not described by the available interpretations of the experimental PES, STS and optical conductivity data. A difficult question of separating the contributions of bandgap charge transport and activated polaron mobility was raised earlier: for example, Gasparov et al. in addition to the bandgap interpretation also discuss the polaronic interpretation of the gaplike feature in optical conductivity~\cite{gasparov2000infrared}.

For nonadiabatic polaron hopping, the energy for reorganization could be expressed as $\lambda = 4E_a$. This energy is required for optical electron transport between sites without moving ions~\cite{austin1969polarons}. 
In optical conductivity and in PES measurements, there is a peak at $\sim0.6~\text{eV}$, which is commonly assumed to be the polaronic peak~\cite{park1998charge,gasparov2000infrared,schrupp2005high}.
Thus, it gives an estimated activation energy for polaron hopping of $E_a = \lambda/4 \sim0.15~\text{eV}$.
Three characteristic energies can be distinguished in the optical conductivity data~\cite{park1998charge,gasparov2000infrared}: two onset energies $E_\text{on1} \sim 0.14~\text{eV}$, $E_\text{on2} \sim 1.0~\text{eV}$ and the first peak energy $E_\text{peak} \sim 0.6~\text{eV}$ (see Figure~\ref{fig:parkscheme}). These values perfectly correspond to our calculated polaron hopping activation energies $E_a$=0.13-0.16~eV, the reorganization energies $\lambda$ 
=0.52-0.64~eV and the $Cc$ bandgap value $E_g$=1.03~eV. We consider this agreement as a support for the proposed polaronic interpretation of the electronic properties of the LT phase.

More studies are required to understand the nature of the experimentally detected signatures of a bandgap at $T>T_V$. Most likely, the current `tiny-gap' experimental interpretations for the HT phase should be changed to a kind of phonon-assisted polaron hopping mechanism. The experiments that were interpreted as a gap closure probably showed some special properties of polaron transport. Our DFT+U results show that changes in orbital ordering can change the band gap between direct and indirect types. Experimental studies of the bandgap type in the LT and HT phases of magnetite might give new interesting information.

The effects of pressure on iron oxides provide new interesting phenomena (see, e.g.,~\cite{zhandun2024orthogonal}). The increase in pressure decreases the temperature of the Verwey transition, and there is some critical pressure around $8~\text{GPa}$ at which the Verwey transition vanishes~\cite{mori2002metallization,rozenberg2006origin,kozlenko2019magnetic}. In addition, it should be mentioned that, for example, in the work of M\^ori et al.~\cite{mori2002metallization} at a pressure of 9 GPa not only the transition disappears, but the resistance increases throughout the temperature range, which clearly indicates that the LT phase shows metallic behavior in contrast to the semiconductor behaviour under normal pressure. Polaron hopping in the LT phase under pressure deserves further study.

\section{Conclusions}

Within the DFT+U model of the magnetite LT phase, we have considered its experimentally proven $Cc$ structure together with several alternatives $P2_1$, $P4_12_12$, $P2/c$, $Pnma$ and $Imma$. The resulting bandgap values lie in the range $E_g$=0.55-1.12~eV with the value for the $Cc$ structure being $E_g$=1.03~eV. 

The trimeron order is found in the $Cc$, $P2/c$ and $Pnma$ structures. For the $Cc$ structure we have reproduced a unique ``bad'' trimeron and have shown that its energy band is located just below the bandgap, which is in reasonable agreement with the site-selective doping detected experimentally for magnetite. Direct bandgaps are found for $Cc$ and $Pnma$ while for other structures considered the bandgaps are indirect.

We have calculated typical electron and hole polaron hopping energies in the LT phase using the $Pnma$ model within the Holstein-Mott theory and the Marcus approach ($E_a$=~0.13~eV and $E_a$=~0.16~eV correspondingly). The model shows that adding a polaron to the LT structure does not have much influence on the trimeron order.

Our results suggest a harmonizing interpretation of the optical conductivity measurements of Park~et~al.~\cite{park1998charge} and Gasparov~et~al.~\cite{gasparov2000infrared}, in which the contributions from small-polaron hopping and bandgap charge transfer are not considered as alternatives, but as two separate mechanisms manifested in optical conductivity spectra at different energies. This interpretation is supported by a remarkable agreement of our DFT+U results and experimental data~\cite{park1998charge,gasparov2000infrared} that can shed new light on the nature of the Verwey transition.

\begin{acknowledgments}
This work was prepared with the support of the Ministry of Science and Higher Education of the Russian Federation (State Assignment No.~075-00270-24-00) and, in part, within the framework of the HSE University Basic Research Program. 
The authors gratefully acknowledge the access to the resources of the Supercomputer Centre of JIHT RAS, the Joint Supercomputer Centre of RAS and the HPC facilities at HSE University~\cite{kostenetskiy2021hpc}.
\end{acknowledgments}


%

\end{document}